# Photochemical Photon Upconverters with Ionic Liquids

*Yoichi Murakami**

Global Edge Institute, Tokyo Institute of Technology, 2-12-1 Ookayama, Meguro-ku,

Tokyo 152-8550 Japan

Email: murakami.y.af@m.titech.ac.jp, Tel/Fax: +81-3-5734-3836

**ABSTRACT:** Photon upconversion based on triplet-triplet annihilation (TTA) of excited triplet molecules is drawing attention due to its applicability for weak incident light, possessing a potential for improving overall efficiencies of solar energy conversion devices. Since energy transfer between triplet levels of different molecules or TTA requires electron exchange between molecules by the Dexter mechanism, inter-molecular collision is necessary and hence the majority of previous studies have been done with volatile organic solvents as fluidic media. This paper presents the development and characterization of a new class of TTA-based photon upconverters with ionic liquids (ILs), which are room temperature molten salts with negligible vapor pressure and high thermal stability. The fabricated samples are found to be stable and the solvation mechanism of the molecules in ILs is proposed. The upconversion quantum yield (UC-QY) is IL dependent and the maximum value observed was approximately 10 % for red-to-blue upconversion. An analytical model was developed to explain the experimental results based on the assumption of efficient energy transfer between molecules. The model agrees with the observed relationship between the excitation intensity and UC-QY, showing that viscosities of the ILs are not a practical problem for the usage of TTA-based photon upconversion.

# 1. INTRODUCTION

Efficient utilization of solar energy has been an important subject. Production of secondary energy such as electrical power (by photovoltaics) and hydrogen (by water-splitting photocatalysts) utilizes only a portion of the solar spectrum for energy conversion. Photons with energies below a threshold energy, which is system dependent, are not utilized in secondary energy generation. To resolve this problem, one proposed strategy has been the use of photon upconversion (UC),[1] which is a process of converting two photons of lower energy into one photon of higher energy. Thus far, UC based on successive absorption of photons by rare-earth elements has been studied for a long time.[2] However, this technology is usually implemented with several orders of magnitude higher light intensities ($10^1$ – $10^4$ W/cm$^2$)[2d–2i] than that of the sunlight (~ 0.1 W/cm$^2$), and UC based on rare-earth elements are currently considered mainly for biological imaging and probing applications.[2f–2i]

An alternative approach is to photochemically perform UC by utilizing triplet-triplet annihilation (TTA) of excited triplet molecules. UC based on this approach was first reported in 1962 with ethanol solution containing two kinds of aromatic molecules,[3] where one of the species absorbs light to sensitize a triplet state and the other species emits blue-shifted photons as a result of TTA; however, the efficiency was limited due to the low triplet-sensitizing efficiency of the molecule. Recently, such triplet-triplet annihilation based photon upconversion (TTA-UC) is being studied using metallated organic molecules (such as metallated porphyrins and phthalocyanines) as efficient triplet sensitizers with molecules possessing high fluorescence quantum yields (such as polycyclic aromatic molecules) as emitters.[4] This strategy has turned out to work with weak non-coherent light sources of intensities even close to terrestrial solar irradiance.[4b,4f] Since the energy transfer between triplet states of the molecules relies on the Dexter mechanism,[5] the fluidic media has been used to allow for sufficient diffusional motion and collisions in order to realize efficient energy transfers between molecules. Due to this reason, the majority of thus far reported TTA-UCs have been performed with organic solvents such as toluene and benzene.[4] However, the use of such flammable and volatile solvents is a hurdle for their practical application.

This paper reports the development and characterizations of a new class of TTA-based photon upconverters fabricated with ionic liquids (ILs) as the fluidic host. Ionic liquids are room-temperature molten salts,[6] which have recently drawn attention partly because of their negligible vapor pressures[6,7] and high thermal stabilities up to several hundred degrees Celsius.[6,8] As imagined by their *ionic* nature, ILs are known to have polarities similar to those of alcohols,[9] while the polycyclic aromatic molecules used in TTA-UC are non-polar or weakly polar. This leads to an intuitive prediction that ILs are an unsuitable media for the purpose of TTA-UC. Contrary to this expectation, however, the molecules have been found to be stable in ILs for a long time. In this paper, the solvation mechanism is proposed and the characterization of developed samples is presented along with the measured upconversion quantum yields (UC-QYs), regarding photon upconversion for the peak-to-peak anti-Stokes energy shift of $\Delta E \sim$ 0.65 eV corresponding to red-to-blue upconversion.

## 2. EXPERIMENTAL SECTION

The ILs tested in this report are listed in Table 1. These were purchased from IoLiTec (#1 – #7, #10), Covalent Associates (#1, #2, #6, #9), Kanto Chemical (#8, #13), Merck (#11), and TCI (#12), and used without further purifications. They were stored under nitrogen until just before use. As the triplet sensitizer and photon emitter molecules, *meso*-Tetraphenyl-tetrabenzoporphine Palladium (PdPh$_4$TBP) and perylene[4f] (**1** and **2** in Figure 1a, from Sigma-Aldrich) were used, respectively.

It has been known that ILs, including those employed in this study, have polarities similar to those of short chain alcohols.[9] On the other hand, polycyclic aromatic molecules used for TTA-UC, including **1** and **2**, are non-polar or weakly polar and generally do not dissolve in methanol. Figure 1b is a photograph taken 24 h after the powders of **1** and **2** were sprinkled over IL #1 held in a quartz mortar. The powders were still floating on the IL's surface, while only a part of the IL around the powders was faintly colored, showing that they hardly dissolve in the IL spontaneously.

**Table 1.** List of the ionic liquids tested. Right column: UC-QYs measured by 632.8 nm CW excitations (30 mW) for the samples fabricated with [S] = 1 × 10$^{-5}$ M and [E] = 3 × 10$^{-3}$ M.

| # | Ionic Liquid | Mixture Uniformity | QY (%) |
|---|---|---|---|
| 1 | [C$_2$mim][NTf$_2$] | Yes | 3.3 |
| 2 | [C$_4$mim][NTf$_2$] | Yes | 4.4 |
| 3 | [C$_6$mim][NTf$_2$] | Yes | 5.2 |
| 4 | [C$_8$mim][NTf$_2$] | Yes | 4.2 |
| 5 | [C$_2$dmim][NTf$_2$] | Yes | < 1 |
| 6 | [C$_3$dmim][NTf$_2$] | Yes | < 1 |
| 7 | [C$_4$dmim][NTf$_2$] | Yes | 10.6 |
| 8 | [NR$_4$][NTf$_2$] | Yes | 2.5 |
| 9 | [C$_3$dmim][CTf$_3$] | Yes | – |
| 10 | [C$_2$mim][CH$_3$CO$_2$] | No | – |
| 11 | [C$_2$mim][CF$_3$CO$_2$] | No | – |
| 12 | [C$_3$mim][I] | No | – |
| 13 | [NR$_4$][BF$_4$] | No | – |

Abbrebiations:
[C$_n$mim]: 1-*Alkyl*-3-methylimidazolium
[C$_n$dmim]: 1-*Alkyl*-2,3-dimethylimidazolium
[NR$_4$]: N,N-Diethyl-N-methyl-N-(2-methoxyethyl)ammonium
[NTf$_2$]: Bis(trifluoromethylsulfonyl)imide
[CTf$_3$]: Tris(trifluoromethylsulfonyl)methide

The sample fabrication procedure developed is described as follows. First, stock solutions of **1** and **2** in toluene (concentrations: 4 × 10$^{-4}$ M and 4 × 10$^{-3}$ M, respectively) were prepared and stored under nitrogen until just before use. The stock solutions were added with a mechanical pipette to an IL held in a glass vial, which resulted in a layer-separation (panel (i) of Figure 1c). Typically, the sensitizer stock (10 – 50 μl) and the emitter stock (100 – 300 μl) were added to an IL (400 μl). The ILs with high hydrophobicity (#1 – #9) have been found to be miscible with toluene in finite amounts (e.g., up to ~ 240 μl toluene miscible with 400 μl of IL #1[10]). For this group of ILs, the solutions were made uniform looking using shear mixing done by gentle repeated suction-and-ejection with a glass Pasture pipette (panel (ii) of Figure 1c). Immediately after this, the vial was capped and underwent moderate ultrasonication for 10 – 20 minutes. Subsequently, the vial was set in a vacuum chamber and pumped by an oil-free scroll pump for 4 – 10 h to remove the toluene (panel (iii) of Figure 1c). In the case of higher molecular densities that could not be achieved in one step due to miscibility limitations, the stock solution was further added at this point, followed by an additional loop of shear mixing, ultrasonication, and vacuum evacuation. The emitter stock solution of 300 μl corresponding to the perylene density of 3 × 10$^{-3}$ M was mixed into the ILs by two separated steps, 200 μl in the first step and remaining 100 μl along with the sensitizer stock solution in the second step. The vial was then set in a purpose-made high-vacuum chamber that was built inside of a stainless steel (SUS) vacuum glovebox. The high-vacuum chamber was evacuated by a turbo-molecular pump for at least 12 h to reach $10^{-4} - 10^{-5}$ Pa, while the vapor pressure of ILs at room temperature are several orders of magnitude lower ($10^{-10} - 10^{-9}$ Pa).[7] Finally, the high-vacuum chamber was opened in the Ar-filled SUS glovebox, and the sample liquid was injected and sealed in an appropriate glass container depending on the purpose. For UC-QY measurements, the sample liquid was injected into a square cross-section quartz tube and sealed with solder inside the Ar-filled glovebox. Details of UC-QY measurements are described in the Supporting Information. As for the rest of the tested ILs shown in Table 1 (#10 – #13), which are moderately-to-highly miscible with water,[6b] they have been found to be unable to form uniform mixtures with the stock solutions, and hence are not further investigated in this paper.

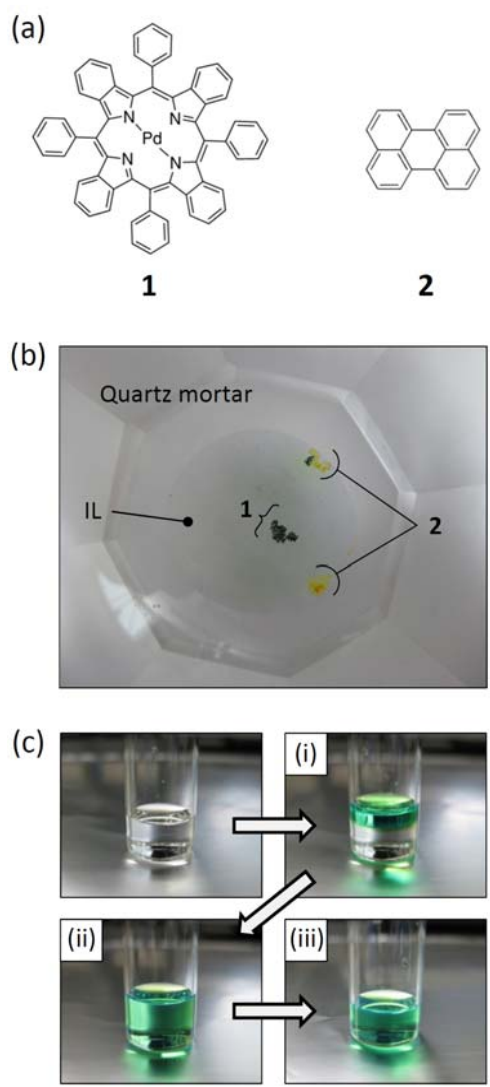

**Figure 1.** (a) The sensitizer (PdPh$_4$TBP, **1**) and the emitter (perylene, **2**) used in this study. (b) A photograph taken 24 h after **1** and **2** were sprinkled over ionic liquid (IL #1) held in the bottom of a quartz mortar. (c) Photographs taken at each step of the sample fabrication.

## 3. RESULTS

Figure 2a shows a photograph of an upconversion of continuous wave (CW) laser light (632.8 nm, 10 mW). Bright blue emission was clearly seen under room light illuminations. Figure 2b shows typical emission spectrum. The full emission spectrum including phosphorescence emission from the sensitizer is shown by Figure S1 in the Supporting Information. As was shown by Figure 1b, the powders of **1** and **2** (which prefer non-polar media) show poor spontaneous dissolution into the ILs (which are polar media). This causes concerns about the stability of these molecules in ILs after the removal of toluene by vacuum pumping. To examine this, an aging experiment (maintained at 80 °C for 100 h followed by at R. T. for 39 h) was performed, as shown in the Supporting Information. Figures S2a – S2d show that the optical absorption spectra of the samples exhibited no change caused by this aging process, indicating that the molecules had been stably solvated in the ILs. Further, the sample emits bright upconverted blue emission 7 months after its fabrication as shown by Figure S2e, demonstrating the temporal stability.

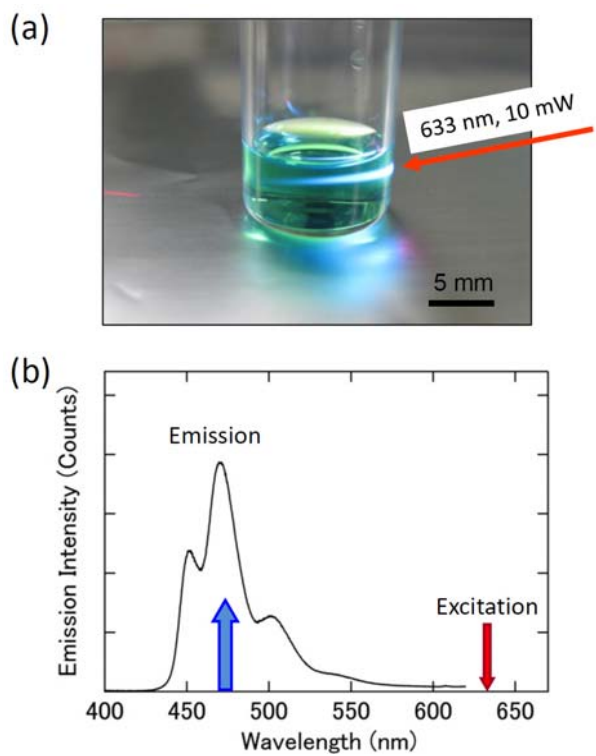

**Figure 2.** (a) A typical looking photograph of an upconversion of 632.8 nm incident CW light to blue emission under room light illuminations by the sample made with IL #1. (b) Photoemission spectrum.

The UC-QYs measured for samples made with ILs #1 – #8 are shown in the right column of Table 1. The excitation light source was a 632.8 nm CW laser (30 mW with a spot diameter of 0.8 mm, corresponding to ~ 6 W/cm$^2$). The reproducibility of the UC-QYs shown in Table 1 has been confirmed by repeated experiments. The variance of the UC-QY for a specific IL was found to be up to 20 %, and this variance may have been caused by untraced impurities or residual O$_2$. Multiple samples of different ILs were simultaneously fabricated and then measured under the same optical alignment, so that the measured UC-QY values were inter-comparable. From this, the order of the UC-QYs for different ILs shown in Table 1 was found not to change. Within the measurement of one identical sample, the uncertainty of the photoemission intensity or UC-QY has been found to vary by up to 5 %. As for sample #6 the UC-QY of 1 % or less has been repeatedly confirmed with ILs purchased from different suppliers (IoLiTec and Covalent Associates).

Along the series of [C$_n$mim][NTf$_2$] (#1 – #4 for $n$ = 2, 4, 6, and 8), the UC-QY takes a maximum at $n$ = 6. Especially among ILs #1 – #3, the result is noteworthy because the ILs with longer alkyl chain, which have higher viscosities,[11] resulted in higher UC-QYs. This is different from previous findings for TTA-UC, where an increase in the media's viscosity led to a significant lowering of UC-QYs.[12] Sample #7 showed the highest UC-QY of approximately 10 % among the samples fabricated. The UC-QY of 6.4 % (after a factor of 2 difference in the definition of UC-QY is corrected) was previously reported for the same sensitizer and emitter dissolved in toluene.[4f] The peak-to-peak anti-Stokes energy shift, Δ$E$, is approximately 0.65 eV in this study. Castellano and co-workers employed boron dipyrromethene chromophores as emitter in benzene and reported the UC-QYs of ~ 6 % (for Δ$E$ ~ 0.4 eV) and ~ 15 % (for Δ$E$ ~ 0.2 eV).[4i,13] Generally, the magnitude of UC-QY decreases as the designed Δ$E$ increases; for example, a recent paper[4j] reported a record shift of Δ$E$ ~ 0.8 eV but the UC-QY was approximately 1 %. Apart from CW excitation, the UC-QY of 16 % for Δ$E$ ~ 0.4 eV has been reported for 1 kHz femto-second laser pulse excitations (the intensity during pulse was as high as 13 GW/cm$^2$, corresponding to a steady-state triplet concentration achieved with 60 W/cm$^2$ CW excitation) of toluene solutions.[14] Table 2 shows the influence of the sensitizer and emitter concentrations on the UC-QY; higher emitter concentration and lower sensitizer concentration resulted in higher QY-UC. The latter may be explained in terms of a quenching of the excited emitter molecules by the ground state sensitizer molecules, as previously proposed.[14]

**Table 2.** Dependence of the UC-QY on the sensitizer and emitter concentrations measured by 632.8 nm CW excitations (27 mW) for the samples fabricated with IL #3.

| Sensitizer (M) | Emitter (M) | QY (%) |
|---|---|---|
| 3 × 10$^{-5}$ | 5 × 10$^{-4}$ | 1.9 |
| 1 × 10$^{-5}$ | 1 × 10$^{-3}$ | 4.3 |
| 3 × 10$^{-5}$ | 1 × 10$^{-3}$ | 3.1 |
| 1 × 10$^{-4}$ | 1 × 10$^{-3}$ | 1.4 |
| 3 × 10$^{-5}$ | 3 × 10$^{-3}$ | 4.4 |

## 4. DISCUSSION

**Data Interpretation.** As has been shown by Figure S2, both the sensitizer and emitter molecules **1** and **2** are stably solvated in the ILs for a long time. This opposes intuitive expectation because ILs are polar solvent[9] while the aromatic molecules prefer non-polar environments. However, this simplified view that is sorely rooted in the consideration of polarity neglects cation-π interaction.[15] I propose cation-π interaction is the mechanism for the observed stabilization of the molecules in the ILs. The emitter and sensitizer molecules are abundant in π electrons. The cation-π interaction has been proposed to be an important stabilization mechanism based on an electrostatic attraction between a positive charge and a π face of an aromatic system.[15] To examine its role in the present study, the miscibility of ILs with benzene ($C_6H_6$) and cyclohexane ($C_6H_{12}$), both typical non-polar solvents, were compared. Figure S3 in the Supporting Information shows that benzene was moderately miscible but cyclohexane, which is similar but lacking π electrons, is completely immiscible with the same ILs. The result implies that an existence of π electrons in the molecules, not the extent of polarity, dominates the solvation in the present study.

The UC-QYs in Table 1 are dependent on IL employed. This variance is considered to be caused by the differences in the microscopic electronic environment (electrostatic interaction, hydrogen bonding, dipole-dipole interaction, etc.) as well as kinetic environment in an IL. This essentially arises from the molecular structures of ions and the spatial coordination of anions around a cation (or vice versa). So far, it has been well established that different molecular structures, or conformational isomers, coexist in the liquid phase of an ionic liquid.[16] The number of possible conformers rapidly increases as the length of an alkyl chain in an imidazolium cation increases.[17] As for a butyl imidazolium cation ($[C_4mim]^+$), the number of possible structures reported by DFT calculations has been between 9 to 12, to the author's best knowledge, depending on previous works.[16d,16f,16g,17] Among these predicted structures, 4 conformers of $[C_4mim]^+$ have been experimentally observed for the liquid phase of $[C_4mim][BF_4]$.[16c] However, quantitative determination of their relative population is still of significant experimental

difficulty.[16f] Recent DFT calculations have shown that the conformational isomerism of an imidazolium cation significantly changes the charge distribution on the imidazolium ring.[16g] Namely, the actual number of molecular structures existing in the liquid phase of ionic liquids, their population ratios, as well as their resultant effects on the microscopic electronic environment within ILs, remain open questions.

A proton at the 2-position of an imidazolium cation ring has been known to play an important role due to the localization of positive charge around this position. The strong interaction of this position with anions has been explained in terms of hydrogen bonding,[18] while the nature of the interaction is still under debate.[19] The methylation of this position (i.e., modification of [$C_n$min]$^+$ into [$C_n$dmin]$^+$) is therefore expected to weaken the anion-cation interaction in general.[20] It has recently been reported that the methylation of this position changes the energetically preferred sites of anions around the cation, the number of possible conformers, and the equilibrium population ratio among different conformers in the liquid phase.[16f,20] The difference in the observed UC-QYs between #1 and #5, or #2 and #7, may originate from such complex influences at the microscopic level.

To investigate further, the dependence of photoemission on the excitation power was examined. Figure 3a shows the dependence of measured UC emission intensity on the excitation power, plotted on a double-logarithmic scale. At lower powers, the slopes of the plots are close to (but less than) two, and as the power increases, the slopes monotonically decrease and approach unity. Accordingly, the UC-QYs start to saturate toward their respective constant values (Figure 3b). In previous TTA-UC studies measured with CW excitation (where volatile organic solvents were employed for media), the emission intensities were reported to change quadrically with the excitation powers.[4g,4i] This quadratic dependence was explained that the TTA-UC process is a two-body annihilation process and hence the chance for a triplet molecule to find an annihilation partner within its life time is proportional to the square of the created triplet concentration. Under such circumstances, the UC-QY increases linearly with the excitation power. Recently, Cheng et al.[14] have reported a quadric-to-linear transition in the photoemission intensity by exciting their toluene-based samples with intense femtosecond laser pulses.

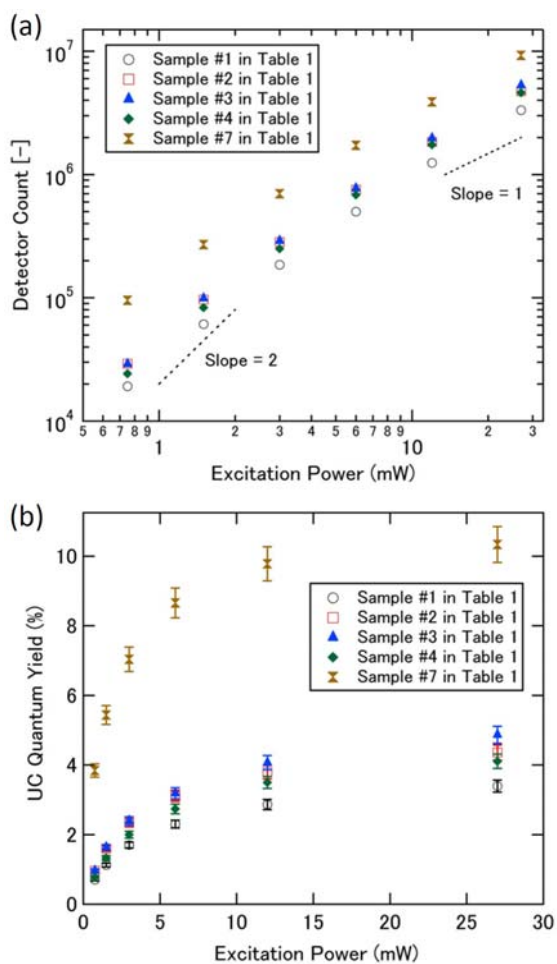

**Figure 3.** (a) Dependence of the upconversion emission intensities on the excitation power of a 632.8 nm CW laser light for the samples fabricated with ILs #1 – #4, and #7, plotted in double logarithmic scale. Dashed lines represent linear and quadric increments shown for eye-guides. (b) Dependence of the UC-QY on the excitation power of the 632.8 nm CW laser light for the same samples. The error bars account for the ±5 % certainty in the photoemission measurement.

The authors have suggested two possibilities for the observed quadric-to-linear transition: (i) photo-bleaching caused in the sensitizing molecules by the intense pulse excitations and (ii) achievement of sufficiently high triplet densities so that virtually all of the created triplets could find their annihilation partners within their lifetimes and hence the emission intensities linearly correlated with the excitation powers. The authors stated that both mechanisms were considered to coexist in their results.[14]

In the present study, the excitation was moderate CW light (0.15 – 6 W/cm$^2$) and hence the first possibility has been excluded. The results in Figure 3 are explained as follows based on the second possibility. When the excitation power is low and hence the triplet population is low, the TTA rate is determined by the probability of triplet species to find annihilation partners. This is a competing process with their decay to the ground state, which may be caused by spontaneous intersystem crossing or by the residual oxygen molecules that efficiently quench excited triplet states at almost diffusion-controlled rate. However, when the excitation intensity is high enough to create sufficient spatial density of the triplets *and when the oxygen concentrations in the media is sufficiently low*, virtually all of the created triplets find annihilation partners before they decay into the ground state even with weak excitations, leading to the saturation of the UC-QY values.

As stated above, the excitation in the present study is CW and is much weaker than the previous study. In addition, the measured viscosities of the ILs are reported to be about two-orders of magnitude higher than the common volatile solvents. In the previous TTA-UC study,[14] however, the authors also mentioned the possibility of the incomplete removal of oxygen molecules from the samples. One distinct difference of the present study from the previous studies done with volatile solvents is the use of IL-based samples that can be thoroughly degassed by pumping with ultra-high vacuum turbo-molecular pumps exploiting negligible vapor pressure of ILs.

**Modeling and Analysis.** Whether or not efficient triplet energy transfer between the molecules is even possible in ILs is examined in the following. The energy level diagram in the present study is shown by Figure 4a. For typical conditions in this study, the emitter concentration is much higher than that of the sensitizer, i.e., [E] >> [S]. In palladium porphyrins the $^1S^* \rightarrow {}^3S^*$ intersystem crossing (ISC) is known to occur with almost unity quantum yield.[21] Under this circumstance and with steady CW excitation, the number conservation of the energy-carrying quanta is described for the sensitizer by the following equation, in terms of the densities of the sensitizer molecules in the lowest triplet state $[^3S^*]$, the emitter molecules in the lowest triplet state $[^3E^*]$, and the emitter molecules in the ground singlet state $[^1E^G]$, all of which are in unit of molar (M)

$$N_{ex} = k_{T(S)}[^3S^*] + k_{TET}[^3S^*][^1E^G] + k_{TTA}[^3S^*][^3E^*]. \qquad (1)$$

In Eq. (1), $N_{ex}$ is the molar rate of photons absorbed by the sensitizer (M s$^{-1}$), $k_{T(S)}$ is the triplet decay rate of the sensitizer (s$^{-1}$), $k_{TET}$ is the triplet energy transfer rate from sensitizer to emitter (M$^{-1}$ s$^{-1}$), and $k_{TTA}$ is the TTA rate between molecules in triplet states (M$^{-1}$ s$^{-1}$), as indicated in Figure 4a. On the other hand, the number conservation equation for the emitter is

$$k_{TET}[^3S^*][^1E^G] = k_{T(E)}[^3E^*] + k_{TTA}[^3S^*][^3E^*] + 2k_{TTA}[^3E^*]^2, \qquad (2)$$

where $k_{T(E)}$ denotes the triplet decay rate of the emitter (s$^{-1}$). From Eqs. (1) and (2), the equation

$$N_{ex} = k_{T(E)}[^3E^*]\left(1 + \frac{k_{T(S)}[^3S^*]}{k_{T(E)}[^3E^*]}\right) + 2k_{TTA}[^3E^*]^2\left(1 + \frac{[^3S^*]}{[^3E^*]}\right) \qquad (3)$$

is obtained. The rate of photon emission from the lowest exited singlet level of the emitter, denoted by $N_{em}$ (M s$^{-1}$), is expressed by the following equation

$$N_{em} = \varepsilon \varphi k_{TTA} [^3E^*]^2, \tag{4}$$

where $\varepsilon$ and $\varphi$ denote the photoemission quantum efficiency from the $^1E^*$ level and the statistical branching ratio of the $^3E^* + {}^3E^* \to {}^1E^*$ in the TTA process, respectively ($0 \leq \varepsilon \leq 1$; $0 \leq \varphi \leq 1$). Further, the UC-QY ($\theta$: $0 \leq \theta \leq 1$) is defined as

$$\theta \equiv \frac{2N_{em}}{N_{ex}} = \frac{2\varepsilon\varphi k_{TTA}[^3E^*]^2}{N_{ex}}, \tag{5}$$

where Eq. (4) was used to derive the right-most term.

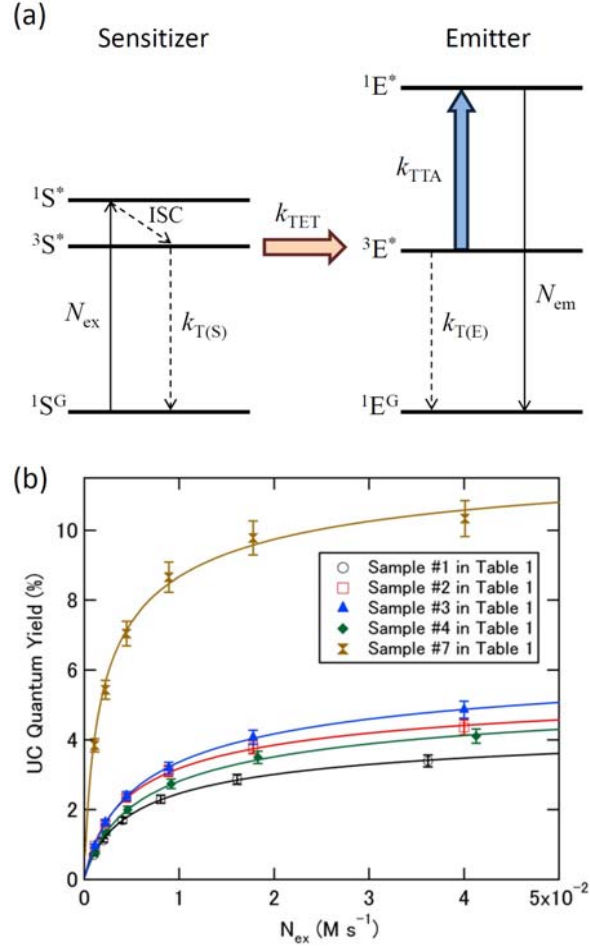

**Figure 4.** (a) Schematic of the energy level diagram in the present system. See the main text for the definitions of the symbols. (b) Plot of the experimental UC-QY vs. $N_{ex}$ relations for the same data set as those shown in Figure 3b and the curves fitted to them by Eq. (12). The error bars account for the ± 5% measurement certainty.

Before proceeding further, an assumption of

$$k_{TET}[^1E^G] \gg k_{T(S)}, \tag{6}$$

which is the condition for a highly efficient donor-acceptor energy transfer,[22] is considered. This assumes that the energies of $^3S^*$ are transferred to the emitter more rapidly than they decay into $^1S^G$. An energy transfer from the sensitizer to the emitter is considered here because the triplet lifetime of the former is much shorter than that of the latter and [E] >> [S]. To see if this holds, the order of $k_{TET}$, which is considered to be similar to the diffusion controlled rate constant, $k_{dif}$, due to its spin-conserving exchange mechanism (Dexter mechanism), is estimated using the following Debye Equation:[22]

$$k_{TET} \approx k_{dif} = \frac{8RT}{3000\eta}. \tag{7}$$

In Eq. (7), $R$: the gas constant ($8.31 \times 10^7$ erg mol$^{-1}$), $T$: temperature (K), and $\eta$: viscosity of the media expressed in unit of poise (P). Since the typical viscosities of the ILs in this study fall in the range of 0.4 – 1 P,[11] the range of $k_{dif}$ calculated by Eq. (7) is $6.6 \times 10^7 – 1.7 \times 10^8$ M$^{-1}$ s$^{-1}$, and here $1 \times 10^8$ M$^{-1}$ s$^{-1}$ is taken as the representative value. Accordingly, since the typical emitter concentration in this study is $3 \times 10^{-3}$ M, $k_{TET}[^1E^G]$ is estimated to be ~ $3 \times 10^5$ s$^{-1}$. On the other hand, the triplet lifetime of PdPh$_4$TBP (**1**) in polar media has been reported to be ~ 260 μs,[23] which corresponds to $k_{T(S)}$ ~ $4 \times 10^3$ s$^{-1}$. Although the $k_{T(S)}$ in ILs may not be the same rigorously speaking, as an order-of-magnitude estimation, $k_{TET}[^1E^G]$ is derived to be two orders of magnitude larger than $k_{T(S)}$, validating the assumption Eq. (6).

In the present study, since [E] is typically two orders of magnitude larger than [S] *and* since the triplet lifetime of the emitter (**2**) in polar media (~ 5 ms)[24] is an order of magnitude larger than that of the sensitizer (**1**) (~ 260 µs),[23] this assumption is to assume that [$^3$E$^*$] >> [$^3$S$^*$]. Based on the same discussion, it is also derived that $k_{T(E)}$ [$^3$E$^*$] is still a few orders of magnitude larger than $k_{T(S)}$ [$^3$S$^*$] despite $k_{T(S)} > k_{T(E)}$. Hence, the order-of-magnitude estimation reduces Eq. (3) to

$$N_{ex} = k_{T(E)}[^3E^*] + 2k_{TTA}[^3E^*]^2. \tag{8}$$

Eq. (8) is readily solved to be

$$\sqrt{2k_{TTA}}[^3E^*] = \frac{1}{2}\left(\sqrt{\frac{k_{T(E)}^2}{2k_{TTA}} + 4N_{ex}} - \frac{k_{T(E)}}{\sqrt{2k_{TTA}}}\right). \tag{9}$$

In the parentheses of Eq. (9), the negative sign has to be chosen because of the physical requisite, [$^3$E$^*$] → 0 as $N_{ex}$ → 0. By substituting Eq. (9) into Eq. (5), the UC-QY is derived as

$$\frac{\theta}{\varepsilon\varphi} = 1 + \frac{k_{T(E)}^2}{4k_{TTA}N_{ex}}\left(1 - \sqrt{1 + \frac{8k_{TTA}N_{ex}}{k_{T(E)}^2}}\right). \tag{10}$$

Here, dimensionless variables Θ and Λ are introduced with a new constant $\alpha$ (in unit of M$^{-1}$ s) as follows

$$\Theta \equiv \frac{\theta}{\varepsilon\varphi}, \quad \Lambda \equiv \alpha N_{ex}, \quad \alpha \equiv \frac{4k_{TTA}}{k_{T(E)}^2}. \tag{11}$$

Finally, Eq. (11) is rewritten in terms of these variables as

$$\Theta = 1 + \frac{1 - \sqrt{1 + 2\Lambda}}{\Lambda} \quad \left(\Leftrightarrow \frac{\theta}{\varepsilon\varphi} = 1 + \frac{1 - \sqrt{1 + 2\alpha N_{ex}}}{\alpha N_{ex}}\right). \tag{12}$$

This dimensionless equation describes the relationship between $\theta$ and $N_{ex}$ in terms of two independent linear scaling factors, $\varepsilon\varphi$ (scaling along the vertical axis) and $\alpha$ (scaling along the horizontal axis). By applying L'Hopital's rule, Eq. (12) has been confirmed to satisfy the physical requisites, "Θ → 0 as Λ → 0" and "Θ → 1 as Λ → ∞". Figure 4b shows the curves by Eq. (12) fitted to the experimental data of $N_{ex}$ vs. $\theta$. Here, the values of $N_{ex}$ were calculated based on the laser beam spot (0.8 mm), the optical path length (1 mm), the molar density of the sensitizer, and the sample's absorbance at the excitation wavelength. The predicted curves (Eq. (12)) show good agreements with the experimental data, corroborating the validity of the assumption (Eq. (6)) in this study. The scaling factors $\varepsilon\varphi$ and $\alpha$ that provide the best fitting are dependent on the IL and are summarized in Table S1 in the Supporting Information.

Finally for this section, the order-of-magnitude estimation performed above for Eq. (6) is conservative, since the actually measured values of diffusion controlled rate constants ($k_{dif}$) in many ILs (including those used in this study) are known to be about an order of magnitude higher than the values estimated using Eq. (7).[25] This also supports the above discussion that ILs are not as viscous of a media as they are generally believed. From the above discussion corroborated by the analysis, it has been shown that ionic liquids, often regarded as viscous media, are not viscous media for the purpose of TTA-UC.

## 5. CONCLUSIONS

Stable photochemical photon upconverters with negligible volatility/flammability have been fabricated using ionic liquids as the fluidic media for triplet energy transfer (Figures 1 and 2). It has been found that the polycyclic aromatic molecules used for the triplet sensitizing and triplet-triplet annihilation are stably solvated in the ionic liquids for a long time (Figure S2). The cation-π interaction has been proposed as the mechanism for the stable solvation. The upconversion quantum yields have been measured for several ionic liquids (Table 1). The maximum upconversion quantum yield observed among the tested conditions was approximately 10 %.

The dependence of the upconversion emission intensities on the excitation power was measured, and it revealed that the upconversion quantum yield saturates to a certain limit depending on the ionic liquid under moderate excitation condition of up to 6 W/cm$^2$ (Figure 3b). This saturation behavior has been explained in terms of efficient energy transfer between the triplet states (Figure 4b). The direct reason for the efficient energy transfer has been attributed to the suppressed oxygen concentrations in the samples, which was made possible with the direct pumping of the samples with an ultra-high vacuum turbo-molecular pump exploiting the negligible vapor pressures of ionic liquids. From the discussion and analysis, it has been found that the triplet energy transfer between the molecules is efficient against decay of the triplet state to the ground state, showing that viscosities of the ionic liquids are not a practical problem for the purpose of triplet-triplet annihilation based upconversions.

In essence, ionic liquids have been found to be useful and are proposed as a media for triplet-triplet annihilation based photon upconverters. The core advantage of using ionic liquids lies in the non-volatility and non-flammability. Although the quantum yields of the developed upconverters are maximized at relatively low incident light powers (~ 3 W/cm$^2$, Figure 3b), for applications to non-concentrated solar irradiances (~ 0.1 W/cm$^2$), upconverters whose quantum yields can reach the maximum at even lower light intensities are desired. This might be achieved by prolonging the triplet lifetime of emitter molecules, if the lifetime is dependent on ionic liquid in which the molecules are solvated. As the number of ionic liquids known thus far exceeds 1,000,000,[6b] further exploration for optimal ionic liquids is expected to give rise to enhancement of the upconversion quantum yield.

**ACKNOWLEDGEMENT**

The author thanks Professors Isao Sato, Osamu Ishitani, Akio Kawai, and Tomokazu Iyoda at Tokyo Institute of Technology for valuable comments. This work was financially supported by MEXT-JST program "Promotion of Environmental Improvement for Independence of Young Researchers" and by KAKENHI (#23686035, Grant-in-Aid for Young Scientists (A)). The findings in this paper have been documented as patent 2010-230938JP and 2011-021136JP.

**Supporting Information Available:** (1) Measurement of the upconversion quantum yield, (2) full emission spectrum from the sample, (3) demonstration of the stability of the samples, (4) miscibility of benzene and cyclohexane with the ionic liquids used, and (5) scaling factors $\varepsilon\varphi$ and $\alpha$ that provided the best fittings.

# Photochemical Photon Upconverters with Ionic Liquids

*Yoichi Murakami**

Global Edge Institute, Tokyo Institute of Technology, 2-12-1 Ookayama, Meguro-ku,

Tokyo 152-8550 Japan

Email: murakami.y.af@m.titech.ac.jp, Tel/Fax: +81-3-5734-3836

**Table of Contents:**

1. Measurement of the upconversion quantum yield

2. Full emission spectrum from the sample

3. Demonstration of the stability of the samples

4. Miscibility of benzene and cyclohexane with the ionic liquids used

5. Scaling factors $\varepsilon\varphi$ and $\alpha$ that provided the best fittings

## 1. Measurement of the upconversion quantum yield

The upconversion quantum yields (UC-QYs) were measured by the following procedure. For this purpose the ILs #1 – #7 provided from IoLiTec were chiefly used, while those provided from Covalent Associates were used mainly for the purpose of checking of the result. Square cross-sectional quartz glass tubes were purchased from VitroCom (QS101, outer and inner dimensions are 2 × 2 mm and 1 × 1 mm respectively) and cut into 25 mm length and washed. Their one end was closed with a burner. After the sample liquids were evacuated by a turbo molecular pump (Pffeifer, HiCube80) for at least 12 h in a high-vacuum chamber installed inside of a SUS vacuum glovebox (UNICO, UN-650F), the chamber was opened in the glovebox under Ar, as described in Section 2. The sample liquids were then injected into the quartz glass tubes to approximately 3/4 of the length using a purpose-made syringe needle, and the tubes' open ends were firmly sealed with Zn/Pb alloy low melting-point solder, all of which were done under the Ar-filled SUS glovebox. Similarly, reference liquid, 9,10-bis(phenylethynyl)anthracene (BPEA) dissolved in toluene ($10^{-5}$ M), was sealed in the same quartz tube.

The liquid-containing quartz tube was rigidly held by a custom-designed stainless-steel mounting block installed on a micrometer-actuated precision XYZ stage, in order to assure the accuracy and reproducibility of the spatial positioning of the tube. The sample and the reference tubes were irradiated with a 632.8 nm CW HeNe laser (Melles Griot, 25-LHP-928) and a 407 nm CW diode laser (World Star Tech, TECBL-30GC-405), respectively. The beam spot diameters at the sample were found to be ~ 0.8 mm using a CCD laser beam profiler (Ophir, SP620). The beam line paths for both 632.8 and 407 nm lasers were carefully aligned and made to be identical using two mechanical irises installed in the optical path. The photoemission was collected in the orthogonal direction to the incident laser beam direction. The emission was collimated by a $CaF_2$ lens and then re-focused by a BK7 glass lens onto the entrance slit of a 30 cm monochromator (PI Acton, SP2300). The spectrum was recorded by a thermoelectrically cooled $1340 \times 100$ pixel Si CCD detector (Princeton Instruments, PIXIS:100BR). The acquired spectrum data was corrected for the wavelength dependences of the grating efficiency and the CCD detector's sensitivity. Along with this, optical absorption spectra of both the sample and the reference, held in a 1 mm thick quartz cuvette (Starna, Type 53/Q/1), were measured using a UV-vis spectrophotometer (Shimadzu, UV-3600).

The values of UC-QY were calculated based on a standard formula with the known emission quantum yield of BPEA (85 %),[S1] the same method as that employed in the previous UC-TTA study.[S2] It is noted that the data analysis in this study has been facilitated by the coincidences in the container material (both quartz) and in the optical path lengths of the photoemission and absorption measurements (both 1 mm). Definition of the UC-QY in this study is based on Eq. (5); the UC-QY is 100 % when all the absorbed photons are converted to upconverted photons.

## 2. Full emission spectrum from the sample

The full emission spectrum from the sample made with IL #7 is shown by Figure S1. The sensitizer and emitter concentrations are $5 \times 10^{-5}$ M and $1 \times 10^{-3}$ M, respectively. In this figure, the wavelength dependences of the grating efficiency and the CCD sensitivity have been corrected. The feature at 800 nm corresponds to phosphorescence emission from the sensitizer.

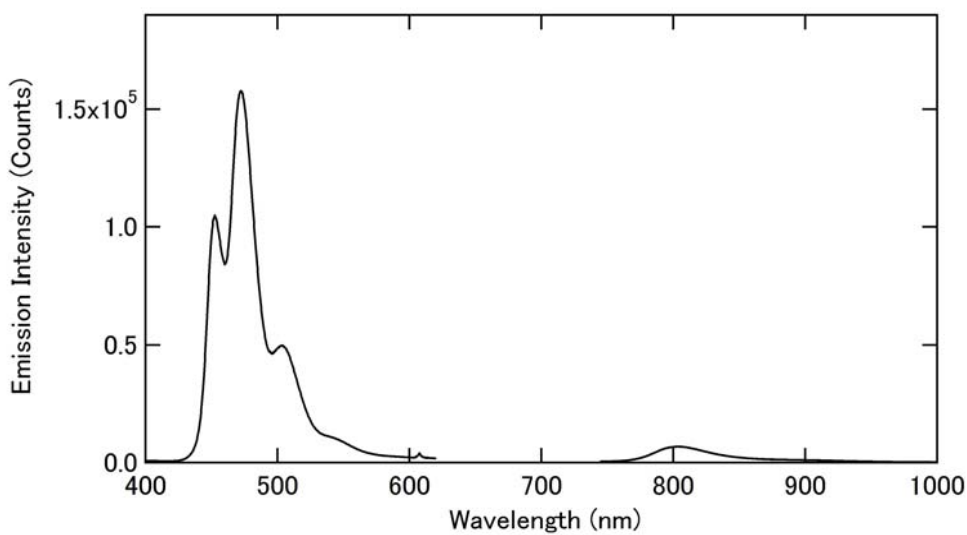

**Figure S1.** Full emission spectrum for 400 to 1000 nm from the sample. Excitation = 632.8 nm, 30 mW.

## 3. Demonstration of the stability of the samples

To check the stability of the samples, the following aging experiment was performed. First, several 1 mm-thick quartz cuvettes were filled with the samples in the Ar-filled glovebox and plugged with the teflon caps. The UV-vis optical absorption spectra of these cuvettes were measured, and then again in the Ar-filled glovebox they were sealed in a purpose-made aluminum chamber. The chamber was then transferred to a forced convection oven and maintained at 80 °C for 100 h, followed by an additional 39 h held at room temperature. As soon as the lid was opened, the UV-vis absorption spectra of the cuvettes were measured. Figures S2a – S2d compare the optical absorption spectra before and after the aging process for the samples made with ILs #1, #2, #6, and #8, respectively. The spectra showed no change by this process, indicating that the molecules had been stably solvated in the ILs. Additional evidence for the temporal stability is shown by Figure S2e, where the sample was left in an atmospheric environment under room light for 7 months since the fabrication. This sample still showed bright blue emission upon CW light irradiation (632.8 nm, 10 mW).

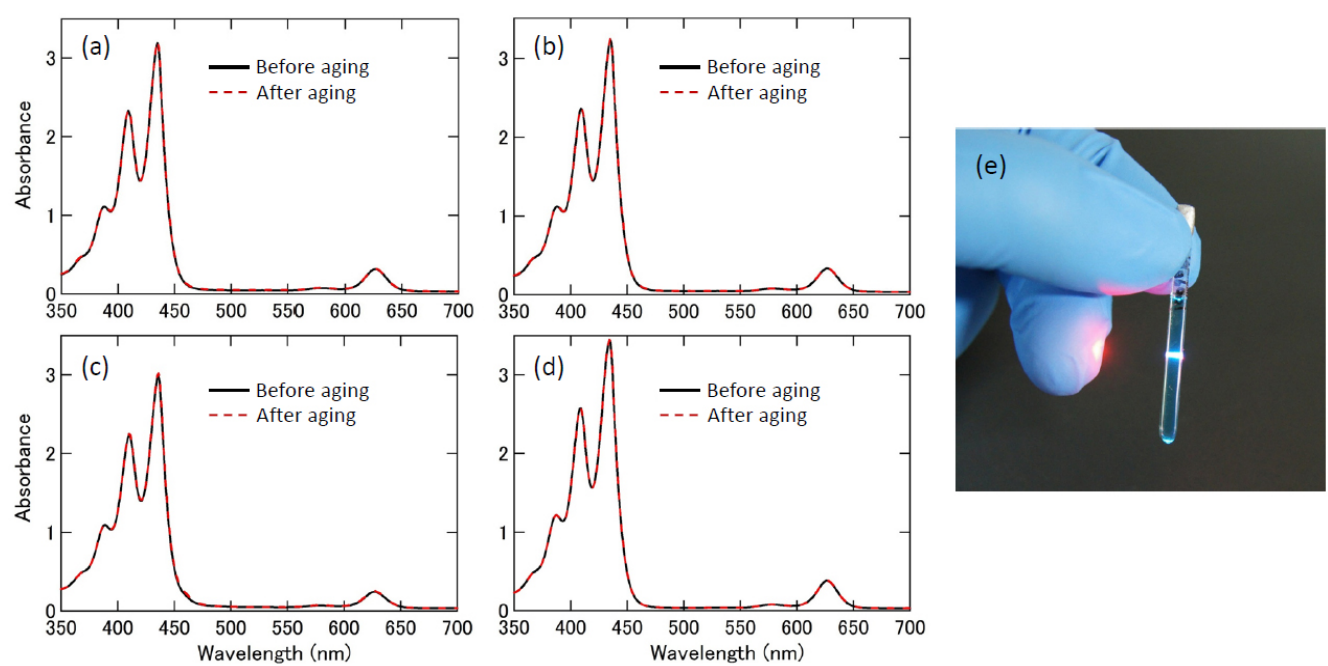

**Figure S2.** (a – d) Comparisons of the optical absorption spectra (path length: 1 mm) taken before and after the aging process for the samples fabricated with (a) IL #1, (b) IL #2, (c) IL #6, and (d) IL #8. The sensitizer and the emitter concentrations were $5 \times 10^{-5}$ M and $1 \times 10^{-3}$ M, respectively. Absorption features in the 350 – 450 nm and 550 – 650 nm ranges are of the emitter and the sensitizer, respectively. The differences in the absorbance magnitude among these spectra were caused by the difficulty in accurately pipetting small volumes of the stock solutions. (e) A photo of an upconversion of 632.8 nm CW laser light (10 mW, ~ 2 W/cm$^2$) taken 7 months after its fabrication. The sample was prepared with IL #1 and had been firmly sealed with low melting point solder in a square cross-section quartz tube (inner: $2 \times 2$ mm, outer: $3 \times 3$ mm, length: 40 mm).

## 4. Miscibility of benzene and cyclohexane with the ionic liquids used

Benzene and cyclohexane were purchased from Wako Chemicals. 300 μl of IL #1, IL #3, IL#7, and IL #8 were held in two sets ($2 \times 4 = 8$) of glass crimp vials, and then sufficient amounts (~ 1 ml) of benzene or cyclohexane were added to each set. The vials were capped with silicone rubber septum using a hand crimper and underwent hand-shaking to mix. The resultant looks of the vials are shown in Figure S3. All of them are separated in two distinct layers, which correspond to the IL that have absorbed and saturated with the organic solvents (bottom layer) and the excess amounts of the organic solvents (upper layer). In the figure, the original surface levels of the IL before adding benzene and cyclohexane are indicated by dotted arrows, and the positions of the interfaces after their addition and mixing are indicated by solid arrows. While benzene is miscible with those ILs by finite amount, cyclohexane is completely immiscible with the ILs. Although both benzene and cyclohexane are common non-polar solvents, there exists a difference in the miscibility with the ILs.

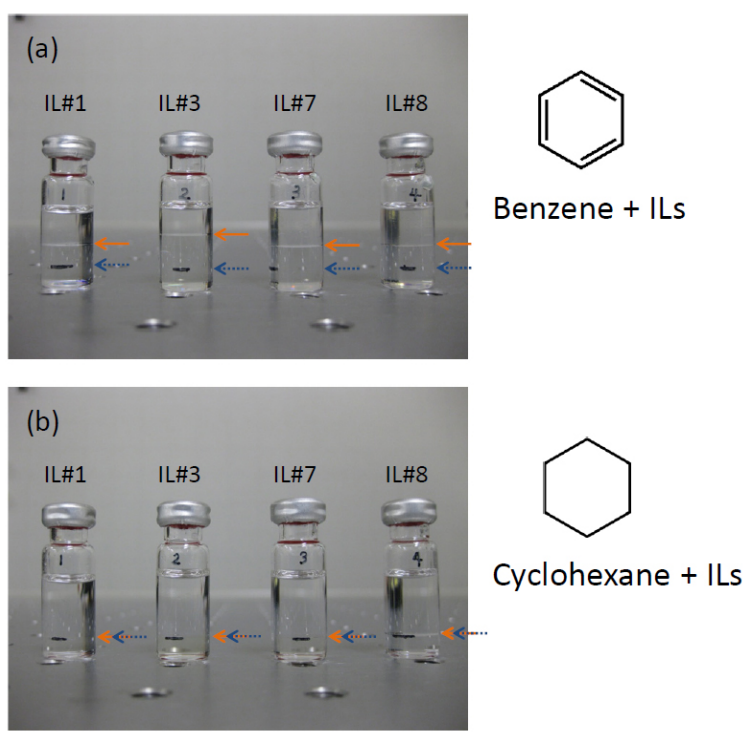

**Figure S3.** Photos of the sample vials taken after mixing of the ILs with (a) benzene and (b) cyclohexane. The dotted arrows indicate the original levels of the ILs in the vials, and the solid arrows indicate the positions of the interfaces between the two layers after mixing.

## 5. Scaling factors $\varepsilon\varphi$ and $\alpha$ that provided the best fittings

The linear scaling factors $\varepsilon\varphi$ and $\alpha$ that provide the best fittings by Eq. (12) in Figure 4b are summarized in Table S1. The magnitude of $\varepsilon\varphi$ sets the upper limit of the UC-QY for each case. Assuming the reported value of $\varepsilon \sim 0.87$ for the case of perylene in polar solvents,[S3] the value of $\varphi$, which is the branching ratio for the singlet formation ($^3E^* + {}^3E^* \rightarrow {}^1E^*$) in TTA process, may have been as high as 0.129/0.87 = 14.8 % for the sample #7. Recently, the singlet formation yield in the TTA process of as high as 33 % was reported for toluene solutions of PQ$_4$Pd and rubrene excited by intense femtosecond laser pulses.[S4] Further, the singlet formation yield in the TTA process under the assumption of an absence of first-order triplet decay in emitter molecules, corresponding to $\varphi$, could be even higher.[S5] Namely, the value of $\varphi$ in TTA-UC could be higher than 1/9, which is the value when the process is purely dictated by the quantum-statistical partitioning condition (i.e., when the formation ratio of singlet, triplet, and quintet is 1:3:5 and all of them are energetically allowed). Therefore, the estimated values of $\varphi$ larger than 1/9 in this study do not violate the previous findings.

**Table S1.** Scaling factors $\varepsilon\varphi$ and $\alpha$ that provided the best fittings by Eq. (12) in Figure 4b.

| # | $\varepsilon\varphi$ ($\times 10^{-2}$) | $\alpha$ ($\times 10^2$ M$^{-1}$ s) |
|---|---|---|
| 1 | 5.0 | 3.9 |
| 2 | 6.2 | 4.4 |
| 3 | 7.1 | 3.4 |
| 4 | 6.1 | 3.1 |
| 7 | 12.9 | 12.4 |